\documentclass[prb,twocolumn,noshowpacs, superscriptaddress]{revtex4}
\usepackage{graphicx}
\usepackage{amsmath}
\usepackage{amsxtra}
\usepackage{amssymb}
\usepackage{dcolumn}
\usepackage{float}
\usepackage{bm}
\usepackage[breaklinks=true,colorlinks,citecolor=blue,linkcolor=blue,urlcolor=blue]{hyperref}

\def\nn{\nonumber}
\def\be{\begin{equation}}
\def\ee{\end{equation}}
\def\bea{\begin{eqnarray}}
\def\eea{\end{eqnarray}}

\begin{document}

\title{\textbf{Dynamical polarizability, screening, and plasmons in one, two, and three dimensional massive Dirac systems} }
\author{Anmol Thakur}
\affiliation{Department of Physics, Indian Institute of Technology Kanpur, Kanpur 208016, India}
\author{Rashi Sachdeva}
\affiliation{Quantum Systems Unit, Okinawa Institute of Science and Technology Graduate University, Okinawa 904-0495, Japan}
\author{Amit Agarwal}
\email{amitag@iitk.ac.in}
\affiliation{Department of Physics, Indian Institute of Technology Kanpur, Kanpur 208016, India}

\date{\today}

\begin{abstract}
We study the density-density response function of a collection of charged massive Dirac particles and present analytical expressions for the dynamical polarization function in one, two and three dimensions.  The polarization function is then used to find the dispersion of the plasmon modes, and electrostatic screening of Coulomb interactions within the random phase approximation. 
We find that for massive Dirac systems, the oscillating screened potential (or density) decays as  $r^{-2}$ and $r^{-3}$ in two, and three dimensions respectively, and as $r^{-1}$ for one dimensional non-interacting systems. 
However for massless Dirac systems there is no electrostatic screening or Friedel oscillation in one dimension, and the oscillating screened potential decays as $r^{-3}$ and $r^{-4}$, in two and three dimensions respectively. 
Our analytical results for the polarization function will be useful for exploring the physics of massive and massless Dirac electrons in different experimental systems with varying dimensionality.
 \end{abstract}

\pacs{}
\maketitle
  
\section{Introduction}
The dynamical polarization function and the associated particle-hole (p-h) excitation spectrum is one of the most  important fundamental quantities for understanding quantum many body effects in fermionic systems. It leads to an understanding of the collective density and spin excitations of the system, the electric polarizability of a material, and the screening properties of the electron gas along with Friedel oscillations, RKKY interaction etc [\onlinecite{Pines, Ando, Giuliani_and_Vignale,  Maier, QP, ag1, ag2, Brey, Schliemann,Schliemann2, Schliemann3}]. The collective density excitations or charge plasmons, for example, has given rise to the exciting field of plasmonics  which merges photonics and electronics at nanoscale dimensions [\onlinecite{Ozbay,marinica, nina}]. 
More recently,  there has been huge interest in plasmons of Dirac materials and particularly in graphene  as it offers a tunable plasmon spectrum via electrostatic control of its carrier concentration, and higher plasmon lifetimes due to high mobility [\onlinecite{GTP, Polini_rev, Stauber,koppens,abajo}]. { Conversely, there there have also been efforts to investigate Dirac like excitations in plasmonic materials [\onlinecite{weick1,weick2}]}. 

The polarization function for regular Schr\"odinger systems with parabolic dispersion is well known in one {(1d)}, two {(2d)}, and three {(3d)} dimensions [\onlinecite{Pines, Ando, Giuliani_and_Vignale}]. For systems with massless Dirac like dispersion in 2d and 3d, the polarization function and charge plasmons has been studied in the context of graphene [\onlinecite{Guinea_NJP2007, SDS1, Polini1, Jablan, AKA}],  topological insulators [\onlinecite{RaghuPRL2009, zhang_ijmp_2013, Pietro}], and  Weyl semimetals [\onlinecite{xiao_arxiv_2014, Panifilov, Hofmann}]. 
Additionally the polarization function for two dimensional massive (gapped) Dirac systems [\onlinecite{Pyatkovskiy, Alireza}] has also been investigated in the context of  buckled honeycomb structures such as silicene [\onlinecite{Tabert, Chang, Duppen}]. The current-current susceptibility for a two dimensional massive Dirac system has been studied in Ref.~[\onlinecite{Scholz}].

In a recent paper [\onlinecite{rashi}], we studied the long wavelength limit of the polarization function and the plasmon dispersion for gapped Dirac systems in all three dimensions, generalizing and validating a similar study for gapless Dirac systems [\onlinecite{SDS2}]. The aim of the present article is to go beyond the long wavelength approximation for massive Dirac systems to calculate the exact polarization function and use it to study charge screening and plasmons in all three dimensions. 
{ This allows us to calculate the exact plasmon dispersion for a given form of interaction, and the scaling of the Friedel oscillations due to a charged impurity at large distances. We find that compared to the plasmon dispersion in the long wavelength limit, the exact plasmon dispersion gets damped generally at a higher 
wave-vector.  Additionally using the exact polarization function in the static limit, we find  that 
the Friedel oscillations in a massive Dirac system decay as $r^{-2}$ and $r^{-3}$ in 2d and 3d respectively similar to the case of parabolic dispersion and unlike the massless Dirac case where the corresponding Firedel oscillations decay as $r^{-3}$ and $r^{-4}$.}

The paper is organized as follows. In Sec. \ref{polfuncsec} we introduce the general form of the dynamic polarization function for massive and massless Dirac systems, and the random phase approximation (RPA)  approach for calculating dispersion of the collective density excitations. In Secs. \ref{polfunc1d}, \ref{polfunc2d} and \ref{polfunc3d}, we present the analytical expressions for the dynamic polarization function of massive Dirac materials, its static limit and Friedel oscillations, charge plasmons, and the gapless limit, in 1d, 2d and 3d respectively. { Finally,  we discuss the non-relativistic limit of the polarization function in Sec. \ref{NR}, and  summarize our findings in Sec.~\ref{summary}.}

\section{Polarization function}
\label{polfuncsec}

We consider a  massive (gapped) Dirac plasma in {1d}, {2d}, {3d} and   analytically calculate the finite frequency and finite wave-vector density-density response function.
The polarization function for a massive Dirac material is given by [\onlinecite{Pyatkovskiy, rashi}]
\be \label{Lindhard}
\Pi(q,\omega)=\frac{g}{L^d} \sum_{{\bf k}, \lambda,\lambda'}F_{\lambda,\lambda'}({\bf k}, {\bf k'})~ \frac{n_{\rm F}(\lambda E_{\textbf{k}})-n_{\rm F}(\lambda'E_{{\bf k'}})}{\hbar\omega + \lambda E_{\textbf{k}} -  \lambda'E_{ {\bf k'}} + i\eta }~,
\ee
where { $L^d$ is the system volume for a $d$ dimensional system, $\omega$ is the angular frequency, $q$ is the wave vector, $\eta \to 0$, $g= g_s g_v$ is the degeneracy due to spin ($g_s$) and valley ($g_v$) degree of freedom. In Eq.~\eqref{Lindhard}}, ${\bf k'} = {\bf k} + \bf {q}$, $\lambda, \lambda' = \pm 1$ denotes the conduction (particle) and valence (hole) bands, 
$E_{\textbf{k}}= \hbar v_{\rm F} \sqrt{k^2+(\Delta/\hbar v_{\rm F})^2}$  with $2 \Delta$ being the energy gap, 
$n_{\rm F} (x)$ is the Fermi function  and $2 F_{\lambda,\lambda'}({\bf k}, {\bf k'}) =1+\lambda\lambda' [\textbf{k}\cdot \textbf{k}'+\tilde{\Delta}^2]/(\tilde{E}_\textbf{k} \tilde{E}_{\textbf{k}'})$ is 
the overlap function. Note that we have defined $\tilde{x} \equiv x/\hbar v_{\rm F}$, which will also be used later in the manuscript. The factor  $g \equiv g_s g_v$ is the degeneracy factor comprising of the  spin degeneracy factor $g_s~(=2)$, and the valley (or pseudo spin) degeneracy factor ({\it e.g.} $g_v = 2$ for graphene and other Dirac materials with honeycomb lattice structure). Given the general relation $\Pi(q, -\omega) = \Pi(q, \omega)^{*}$ as evident from Eq.~\eqref{Lindhard}, and the fact that the polarization function 
depends only on the absolute value of the Fermi energy $\mu $,  we present results only for $\mu> 0$ and $\omega >0$. Furthermore, we work at zero temperature so that the Fermi
functions  can be replaced by Heaviside theta functions, {\it i.e.}, $n_{\rm F}(x) = \theta(\mu-x)$.

Depending upon the placement of the Fermi energy $\mu $, with respect to the energy dispersion, we can split our polarization  function into two parts, namely 
the intrinsic  ($ \mu < \Delta$) and extrinsic polarization ($\mu  > \Delta$). More explicitly, Eq.~(\ref{Lindhard}) can be expressed as
\bea \label{totalpol}
 \Pi~(q,\omega)
=-\chi_{\infty}^-(q,\omega)+ \chi_{\mu }^-(q,\omega)+ \chi_{\mu}^+(q,\omega)~, 
\eea
where $ \Pi_0(q,\omega) \equiv -\chi_{\infty}^-(q,\omega)$ denotes the intrinsic ($\mu < \Delta $) part of the polarization function, and 
$ \Pi_1(q,\omega) \equiv  \chi_{\mu }^-(q,\omega)+ \chi_{\mu}^+(q,\omega) $ denotes the extrinsic part of the polarization function. In Eq.~\eqref{totalpol}, we have defined 
 \bea \label{split}
 & &\chi_{D}^\pm(q,\omega)=-\frac{g}{(2\pi)^d} \int d^d
k ~\theta(D^2-\Delta^2-k^2) \\
& & \times\left(1\pm \frac{\textbf{k} \cdot \textbf{k}'+\tilde{\Delta}^2}{\tilde{E}_{\textbf{k}}~ \tilde{E}_{\textbf{k}'}}\right) \left[\frac{E_{\textbf{k}}\mp E_{\textbf{k}'}}{(\hbar\omega+i\eta)^2-({E}_{\textbf{k}}\mp {E}_{\textbf{k}'})^2}\right]~.\nonumber
\eea
Here the upper and lower signs correspond to intraband and interband  transitions respectively and the parameter $D$ defines the integration limits via the Heaviside $\theta$ function. 
To proceed further for calculating the polarization function, we split the three components of the polarization function in Eq.~\eqref{totalpol}, into the real and imaginary parts using the {  real line version of the Sokhotski-Plemelj theorem}: 
$\frac{1}{x \pm i y}={\cal P}(\frac{1}{x}) \mp i \pi \delta(y)$ 
and then do the wave-vector integration.

Once the non-interacting density-density response function is known, the collective density excitations (plasmon modes) are given by the poles of the interacting density-density response function. Within the { RPA}, the poles of the interacting response function,  coincide with the zeros of the complex longitudinal `dielectric function' $\epsilon(q,\omega)$, which is given by 
\be \label{eq:condition}
\epsilon(q, \omega) = 1 - V_q \Pi(q, \omega) =  0~, 
\ee
where $V_q$ is the Fourier transform of the relevant bare electron-electron (or Coulomb) interaction, and $\Pi(q, \omega)$ is the non-interacting polarizability of the system, given by Eq.~\eqref{Lindhard}. 
The Fourier transform of the { unscreened} Coulomb interaction $V (r) = e^2/(\kappa r)$, in the appropriate $d$-dimensional space is given by 
\begin{subequations} \label{vq}
\begin{align}
      V_q &= \frac{4 \pi e^2}{\kappa q^2} \quad \quad d = 3~,\label{vq3d}  \\
       & =  \frac{2 \pi e^2}{\kappa q} \quad \quad d = 2~,\label{vq2d} \\
	& =  \frac{2 e^2}{\kappa}K_{0}(qa) \quad   d = 1~,\label{vq1d}
\end{align}
\end{subequations}
where $\kappa$ is the background material dependent dielectric constant, and $K_0$ denotes the zeroth order modified Bessel
function of the second kind. Note that in 1d, the length scale `$a$' characterizes   the lateral confinement size 
(say radius of the 1d ribbon), and $V_q\approx -2 e^2 \ln(qa)/\kappa$ for $qa \ll1$ [\onlinecite{1D1}]. We note here that an alternate form of the 1d effective Coulomb potential, as specified in Ref.~[\onlinecite{Giuliani_and_Vignale}], can also be used for the 1D Coulomb potential, and it also yields the same asymptotic form as { Eq.~\eqref{vq1d}} in the $qa\ll1$ limit. 

%, while $V_q \approx e^2/(\kappa q^2 a^2)$ for $q a \gg 1$.  

The static limit of the polarization function is also useful for determining the screened Coulomb potential of a charged impurity.  Consider 
 a localized charged impurity which has a  vacuum potential $\phi_{\rm ext}({\bf r})$ in real space or $\phi_{\rm ext}({\bf q})$ in the momentum space. The potential of the charged impurity,  when it  is embedded  at the origin in the electron liquid,  gets modified into 
\be \label{fdpotential}
\phi ({\bf r}) = \int d{\bf q} ~e^{i {\bf q}\cdot {\bf r}} \frac{\phi_{\rm ext}({\bf q})}{\epsilon({\bf q},0)}~e^{-qs} ~,
\ee
where $s \to 0$.   
Now in the asymptotic limit of $r \to \infty$, the integrand in Eq.~\eqref{fdpotential} is highly oscillatory for any wave-vector, and  thus the 
asymptotic behaviour of the screened potential is determined by the non-analyticity of the static dielectric function, generally at $q=2 k_{\rm F}$, which leads to  Friedel oscillations in the electron gas with a power law decay [\onlinecite{Giuliani_and_Vignale, SimionGiuliani, Chenraikh, Smbadlayan}]. 
A similar behaviour is  also reflected in the charge density, and has been demonstrated experimentally [\onlinecite{expt1, expt2, expt3, expt4}] .  

{  In a realistic experimental situation, in place of Eq.~\eqref{vq}, a screened Coulomb interaction may have to be considered. Any such screened potential will change the Friedel oscillations quantitatively while keeping their qualitative nature intact. However the plasmon dispersion in presence of a screened potential is likely to change qualitatively (in terms of the long wavelength $q$ dependence) as well as quantitatively. }

In the next few sections we calculate the polarization function, long wavelength plasmon dispersion, static dielectric function and Friedel oscillations for 1d, 2d and 3d massive Dirac system.
% one- two- and three- dimensional massive Dirac systems. 
Note that till now, we have explicitly written all the quantities in their usual form with their respective dimensions. However, in what follows we transform all quantities with dimension of energy, {\it i.e.}, $\hbar \omega,~\Delta, ~\mu$ etc, to have the dimension of $1/L$ by dividing them with $\hbar v_{\rm F}$. Thus, we define the following equivalent variables: ${\tilde\omega}  \equiv \omega/v_{\rm F}$, $\tilde{\Delta} \equiv \Delta/(\hbar v_{\rm F})$, ${\tilde \mu} \equiv \mu/({\hbar v_{\rm F}})$ and ${\tilde{E}}_{\bf k} \equiv E_{\bf k}/({\hbar v_{\rm F}})$, with the dimension of a wave-vector.  

%%%%%%%%%%% Polarisation function in 1D%%%%%%%%%%%%%%%%%%%%%%%%%%%%%%%%%%%%%%%%%%
\section{Polarization function in 1d massive Dirac materials}\label{polfunc1d}
In this section, we calculate and present the full analytical expression of the dynamical polarization function for $1$d massive
Dirac materials. We treat the intrinsic and extrinsic components of the polarization function, separately as specified by  Eq.~(\ref{totalpol}), and determine the real and imaginary parts of each component using the Dirac identity. 
As a consistency check, we note that the explicit analytical forms of the real and imaginary part of the polarization function presented below, have been verified against direct numerical integration of Eqs.~\eqref{totalpol}-\eqref{split}.

Let us first consider the imaginary part of the polarization function. 
Performing the wave-vector integration in Eq.~\eqref{split}  and introducing the following notation to express our results in a compact form,  

\bea \label{x0}
x_0 &\equiv& \sqrt{1-\frac{4{\tilde \Delta}^2}{{\tilde \omega}^2-q^2}}~, \nn \\
\alpha(q,{\tilde\omega})&\equiv&\frac{{\tilde \Delta}^2 q^2}{x_0|{\tilde \omega}^2-q^2|^2}~.
%k_1&\equiv&\frac{{\tilde \omega} x_0 -q}{2}~,\nn \\
%k_2&\equiv&\frac{-{\tilde \omega} x_0 -q}{2}~,\nn \\
%\alpha_1(k)&\equiv&\frac{{\tilde \omega} {\tilde E}_{\textbf{k}} -kq-2{\tilde E}_{\textbf{k}}^2}{|{\tilde \omega} k + q {\tilde E}_{\textbf{k}}|}~,\nn \\
%\alpha_2(k)&\equiv&\frac{-{\tilde \omega} {\tilde E}_{\textbf{k}} +kq+2{\tilde E}_{\textbf{k}}^2}{|{\tilde \omega} k + q {\tilde E}_{\textbf{k}}|}~,\nn \\
%\alpha_3(k)&\equiv&\frac{{\tilde \omega} {\tilde E}_{\textbf{k}} +kq+2{\tilde E}_{\textbf{k}}^2}{|{\tilde \omega} k - q {\tilde E}_{\textbf{k}}|}~,
\eea
%%%%%% Regionwise Imaginary%%%%%%%%%%%%%%%%%%%%%%%%
we obtain the imaginary part of the intrinsic polarization function to be 
\be \label{eq:pi01d}
\Im m~\Pi_0^{(1d)}(q, {\tilde \omega})=-\frac{2g\alpha(q, {\tilde \omega})}{\hbar v_{\rm F}}~\theta(\tilde{\omega}^2-q^2-4\tilde{\Delta}^2)~.
\ee
As a consistency check we note that Eq.~\eqref{eq:pi01d} is consistent with the imaginary part of the 
polarization function derived in the context of the Schwinger model in 1+1 dimensions in Ref.~[\onlinecite{Pilon}].

\begin{figure}[t] 
\begin{center} 
\includegraphics[width=0.95\linewidth]{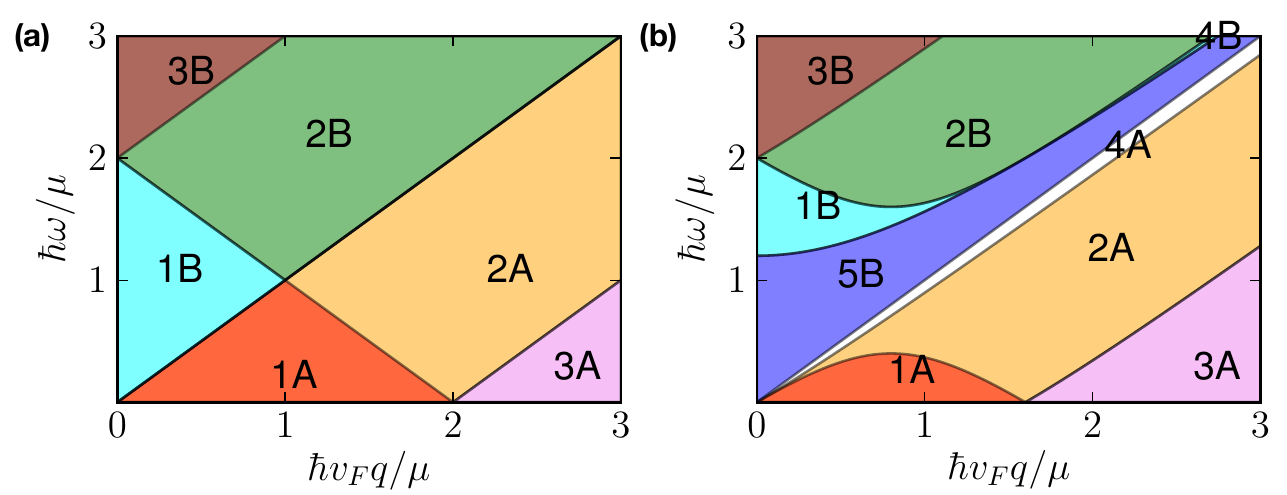} 
\end{center} 
\caption{Different regions in the ${\tilde \omega} -q$ plane used to define the polarization function for (a) massless Dirac systems  and (b) massive Dirac systems, in all three dimensions [see Eq.~\eqref{eq:regions}].  Here $\Delta=0.6 \mu$. The 4B region in panel b), is a minuscule region enclosed by the upper boundary of 5B region and the lower boundary of the 2B region, whose visibility depends on $\Delta$. 
\label{fig1} } 
\end{figure}

A similar calculation for the imaginary part of the extrinsic polarization function yields, 
\be \label{eq8}
\Im m~\Pi_1^{(1d)}=\frac{g}{ {\hbar v_{\rm F}}}
\begin{cases}
-\alpha(q, {\tilde \omega}),  & \text{2A} \\
\alpha(q, {\tilde \omega}),  & \text{2B} \\
2\alpha(q, {\tilde \omega}), & \text{1B} \\
%\alpha(q, {\tilde \omega}), & \text{2B} \\
0,& \text{1A}, \text{3A}, \text{3B}, \text{4A}, \text{4B}, \text{5B}.\\
%0,& \text{4B}, \text{5B}.   
\end{cases}~
\ee
In Eq.~ \eqref{eq8}, we have defined the following regions in the (${\tilde \omega},q$) plane [see Fig. \ref{fig1}]:
\bea \label{eq:regions}
\text{1A} : & &0<{\tilde \omega} < {\tilde \mu}- \gamma(k_{\rm F}-q) ,\nonumber\\
\text{2A} : & &\pm {\tilde \mu} \mp \gamma(k_{\rm F}-q) < {\tilde \omega}  < -{\tilde \mu} +\gamma(k_{\rm F}+q),\nn\\
\text{3A}: & &{\tilde \omega} < -{\tilde \mu}+\sqrt{(q-k_{\rm F})^2+{\tilde \Delta}^2} \nn\\
\text{4A}: & &-{\tilde \mu} + \gamma(k_{\rm F}+ q) <{\tilde \omega} <q \nn \\
\text{1B} : & &q<2k_{\rm F}~ \&~\sqrt{q^2 + 4{\tilde \Delta}^2}<{\tilde \omega} <{\tilde \mu} + \gamma(k_{\rm F}-q), \nonumber \\
\text{2B} :& & {\tilde \mu}+\gamma(k_{\rm F}-q) < {\tilde \omega}  < {\tilde \mu} +\gamma(k_{\rm F}+q),\nn \\
\text{3B} :& & {\tilde \omega}>{\tilde \mu}+\gamma(k_{\rm F}+q)\nn \\
\text{4B} :& & q > 2k_{\rm F}~ \&~ \sqrt{q^2+4{\tilde \Delta}^2} < {\tilde \omega} <{\tilde \mu} + \gamma(k_{\rm F}-q),\nn \\
\text{5B} :& & q < {\tilde \omega} < \sqrt{q^2+4{\tilde \Delta}^2}
\eea
where $ k_{\rm F}=\sqrt{{\tilde \mu}^2-{\tilde \Delta}^2}$ is the Fermi wave-vector, and we have defined the function $\gamma(x) \equiv \sqrt{x^2 + {\tilde \Delta}^2}$. {  Here regions 1A and 2A, denote the regions of the $\omega-q$ space in which only the intraband single p-h transitions are allowed, and  region 2B denotes the regions where only inter-band single p-h transitions are allowed. In all other regions there are no single p-h excitations. Of these the region 1A is very interesting because in 1d, there are no single p-h excitations in the 1A region as a consequence of the restricted phase space for low energy ($\omega \to 0$, and finite $q$) scattering in 1d. In fact it is this absence of low energy finite $q$ single particle excitation in 1d, which leads to the Luttinger liquid behaviour. }
%The different regions  defined above denotes either the presence or absence of the imaginary contribution of the polarization function. Finite imaginary contribution points out that the plasmon modes get damped, and we observe this effect in our calculations further. }
%
%\iffalse
%\fi

For calculating the real part of the polarization function we rewrite the integral in Eq.~\eqref{split} as 
\be \label{eq:11}
\Re e ~\chi_{D}^{\pm}=-\frac{g}{\pi}\int_{- \beta }^{\beta} dk~F_{\pm}(k,k') 
\left(\frac{E_{\textbf{k}}\mp E_{\textbf{k}'}}{{\omega}^2-(E_{\textbf{k}}\mp E_{\textbf{k}'})^2}\right)~,
\ee
where $\beta \equiv \sqrt{D^2-{\tilde \Delta}^2}$.

Performing the integration in Eq.~\eqref{eq:11} we obtain the real component of the intrinsic polarization function in 1d to be,
\be \label{chi1Dr2}
\Re e~\Pi_0^{(1d)}={\frac{g}{2\pi {\hbar v_{\rm F}}}h_1(q,{\tilde \omega})\Big[({\tilde \omega}^2-q^2)+g_1(q,{\tilde \omega})\Big]}~, \\  
\ee
where we have defined the following functions
\bea
h_1(q,{\tilde \omega})&\equiv&\frac{2q^2}{({\tilde \omega}^2-q^2)^2}~, \\
{g_1(q,\tilde{\omega})}&\equiv& \frac{2{\tilde \Delta}^2}{x_0}
\begin{cases}
 \log\left[\left|\frac{(1+x_0)}{(1-x_0)}\right|\right]~, &\text{1A,2A,3A,4A} \\
& \text{1B,2B,3B,4B,} \\
 \log\left[\frac{(1+x_0)}{(1-x_0)}\right]-i\pi~, & \text{5B}.  \nn \\
\end{cases}
\eea
Note that in the 5B region, $x_0$ is imaginary ($x_0 = i |x_0|$), and thus $g_1 (q,\tilde{\omega})$ is real valued. 
Additionally we have also checked numerically that $\Pi^{0}(q,\omega)$ specified by Eqs.~\eqref{eq:pi01d} and \eqref{chi1Dr2} satisfy the 
Kramers-Kronig relations. 

The real component of the extrinsic part of the polarization function is given by 
\bea \label{chi1Dr1}
\Re e~\Pi_1^{(1d)}&=&-\frac{g}{2\pi {\hbar v_{\rm F}}}\Big[f_1\theta({\tilde \omega}^2-q^2-4{\tilde \Delta}^2) \nn \\
&+&f_2\theta(q^2+4{\tilde \Delta}^2-{\tilde \omega}^2)\theta({\tilde \omega}-q)+f_3\theta(q-{\tilde \omega})\Big],\nn\\
\eea
where we have defined the following functions, 
\bea
f_1&\equiv&{\frac{h_1{\tilde \Delta}^2}{x_0}\left[\log\left| \frac{a_1(k_{\rm F})a_1(k_{\rm F}-q)b_4(k_{\rm F})b_3(-k_{\rm F})}{a_1(-k_{\rm F}-q)b_3(k_{\rm F})a_1(-k_{\rm F})b_4(-k_{\rm F})}\right|\right]},\nn\\
f_2&\equiv&{\frac{h_1{\tilde \Delta}^2}{x_0}~i~\text{Im} \left[\log c(k_{\rm F})-\log c(-k_{\rm F})\right]},\\
f_3&\equiv&{\frac{h_1{\tilde \Delta}^2}{x_0}\left[\log \left| \frac{a_1(k_{\rm F})a_1(-k_{\rm F}-q)b_3(-k_{\rm F})b_2(-k_{\rm F})}{a_1(-k_{\rm F})b_3(k_{\rm F})a_1(k_{\rm F}-q)b_2(k_{\rm F})}\right|\right]}~,\nn
\eea
which in turn use the following:
\bea
a_1(k)&\equiv&{\tilde \omega} x_0 + 2k + q ~,\nn \\
b_1(k)&\equiv&2{\tilde \Delta}^2 -k({\tilde \omega} x_0+q)-{\tilde E}_{\textbf{k}}({\tilde \omega} + q x_0)~,\nn \\
b_2(k)&\equiv&2{\tilde \Delta}^2 +k({\tilde \omega} x_0-q)-{\tilde E}_{\textbf{k}}({\tilde \omega} - q x_0)~,\nn \\
b_3(k)&\equiv&2{\tilde \Delta}^2 -k(q+{\tilde \omega} x_0)+{\tilde E}_{\textbf{k}}({\tilde \omega} + q x_0)~,\nn \\
b_4(k)&\equiv&2{\tilde \Delta}^2 -k(q-{\tilde \omega} x_0)+{\tilde E}_{\textbf{k}}({\tilde \omega} - q x_0)~, \nn \\
c(k)&\equiv&b_1(k)~b_4(k)~. 
\eea
Note that in the function $f_2$, $x_0$ is imaginary, hence the overall output of $f_2$ is real. 

%\iffalse
\begin{figure}[t] 
\begin{center} 
\includegraphics[width=1.0\linewidth]{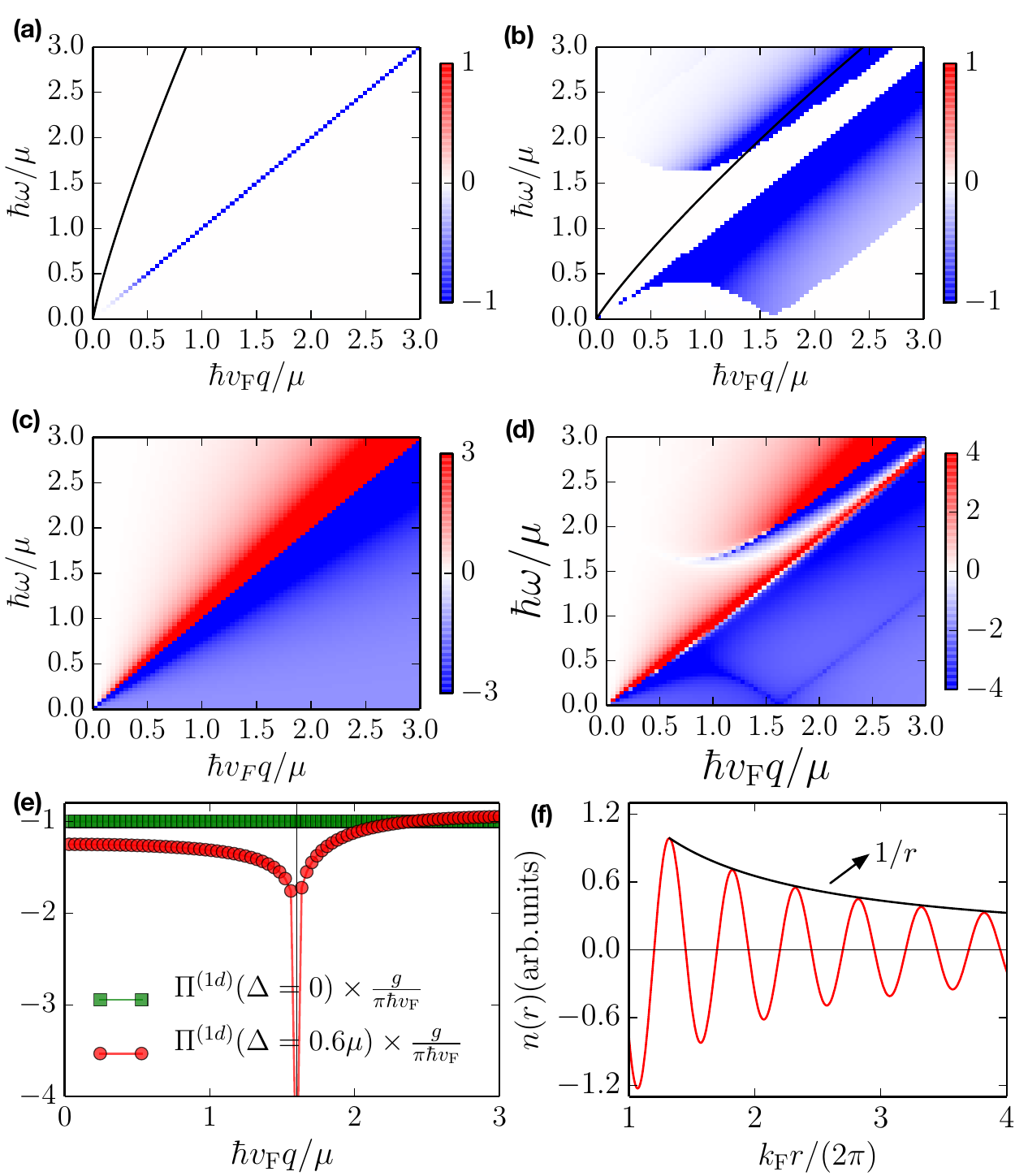} 
\end{center} 
\caption{(a) ${\Im m \Pi(q,{\tilde \omega})}$ in the ${\tilde \omega}-q$ plane,  for a 1d massless Dirac system along with the exact plasmon dispersion --- see Eq.~\eqref{pl_gapless}.   Note that the $q-\omega$ space for p-h excitations is only a straight line due to conservation of chirality and the system being 1d. 
(b) ${\Im m \Pi(q,{\tilde \omega})}$ for a massive 1d Dirac system along with the long wavelength plasmon dispersion --- see Eq.~\eqref{wm1}.   (c) - (d)  ${\Re e \Pi(q,{\tilde \omega})}$ for $\Delta = 0$ and $\Delta = 0.6 \mu$, respectively. In panels (a)-(d) the polarization function is in units of $g/{(\hbar v_{\rm F})}$.  For calculating the plasmon dispersion in panels (a) and (b) we have used $a = 0.02/k_{\rm F}$, and $8 e^2/(\pi \kappa \hbar v_{\rm F}) = 0.6$. 
(e) The static $\Pi(q,{\tilde \omega}=0)$ versus $q$, for the massless and massive Dirac system in $1$d. Note that there is no discontinuity in $\epsilon(q, 0)$ for massless Dirac system in $1$d and as a consequence there are no Friedel oscillations. However in the massive case, there is a logarithmic singularity at $q=2 k_{\rm F}$, which is marked by the thin vertical line.  (f) Asymptotic behaviour of the Friedel oscillations in the density profile in vicinity of a scatterer for the non-interacting case in 1d Dirac system. In general the RPA or the mean field description fails to capture Friedel oscillations in 1d, whose asymptotic behaviour  depends on the strength of electron-electron interaction,  and only in the non-interacting case does one recovers the $r^{-1}$ decay. 
\label{fig2} } 
\end{figure}
%
%\fi
Having obtained the real and imaginary part of the $1$d polarization function, let us consider various limiting cases, 
starting with the static case (${\tilde \omega} \to 0$) in the next subsection. 
\subsection{Static dielectric function and Friedel oscillations}
\label{static1D}
In the static limit, {\it i.e.}, $\omega =0$  case for $1$d massive Dirac systems the intrinsic part of the Lindhard function reduces to 
\be \label{eq:1dstatic0}
\Pi_0^{(1d)}=\frac{-g}{\pi {\hbar v_{\rm F}}}\Bigg[1+\frac{2{\tilde \Delta}^2}{q \sqrt{q^2+4{\tilde \Delta}^2}}\log\left(\frac{\sqrt{q^2+4{\tilde \Delta}^2}-q}{\sqrt{q^2+4{\tilde \Delta}^2}+q}\right)\Bigg],\\
\ee 
and the extrinsic part is given by 
\be \label{eq:1dstatic1}
\Pi_1^{(1d)}=\frac{-g}{\pi {\hbar v_{\rm F}}} \frac{2{\tilde \Delta}^2}{q \sqrt{q^2+4{\tilde \Delta}^2}} \Bigg[\log\left(\frac{(q+2k_{\rm F})\beta_1(k_{\rm F})}{|q-2k_{\rm F}|\beta_1(-k_{\rm F})}\right)\Bigg], \\
\ee
where $\beta_1(k_{\rm F})=k_{\rm F}q+2{\tilde \Delta}^2+\sqrt{(k_{\rm F}^2+{\tilde \Delta}^2)(q^2+4 {\tilde \Delta}^2)}$. Note that $\Pi_1^{(1d)}$ in Eq.~\eqref{eq:1dstatic1} diverges logarithmically as $q \to 2 k_{\rm F}$, similar to the case of 1d parabolic systems \cite{Giuliani_and_Vignale}.
In the limiting case of $q \to 0$, Eqs.~\eqref{eq:1dstatic0}-\eqref{eq:1dstatic1} lead to 
\bea
\Pi^{(1d)}(q\to 0, 0) &=& -\frac{g}{\pi {\hbar v_{\rm F}}} \frac{{\tilde \Delta}}{k_{\rm F}}\left(1+\frac{k_{\rm F}^2}{{\tilde \Delta}^2+{\tilde \Delta} \sqrt{k_{\rm F}^2+{\tilde \Delta}^2}}\right)~\nn \\ 
& = & -g \mu/(\pi {\hbar v_{\rm F}} k_{\rm F})~,
\eea 
 Note that $g \mu/(\pi \hbar v_{\rm F} k_{\rm F})$ is the  density of states of a Dirac system in $1$d at the Fermi energy. 

In general the static dielectric function can also be used to calculate the Friedel oscillations as shown in Appendix \ref{FO}. However in 1d the RPA calculation of the Friedel oscillation breaks down on account of the logarithmic divergence of $\Pi_1^{(1d)}$ in Eq.~\eqref{eq:1dstatic1}. This highlights the failure of RPA or a mean field like calculation to describe a finite wave-vector ($q = 2 k_{\rm F}$) phenomena like the Friedel oscillations for 1d interacting electrons. Interacting 1d electron liquids are more appropriately described by the Luttinger liquid theory which treats the Coulomb interactions in a exact way (barring some anomalous terms). Using Luttinger liquid theory it has been shown that the asymptotic decay of the Friedel oscillation in the density profile with distance from the impurity has a power law decay, with an exponent 
which depends on the interaction strength [\onlinecite{Egger}].  Only in the non-interacting limit, does one recover the $r^{-1}$ decay of the Friedel oscillations as shown in Fig. \ref{fig2}(f).  Generally repulsive interactions tend to slow down the decay of the Friedel oscillations in 1d systems. 
However, for the case of a  gapless system in 1d, with chiral electrons, Friedel Oscillations vanishes completely. This is a direct consequence of the fact that chirality (eigenstate of the { $ {\hat \sigma} \cdot {\bf k}$} operator), in a gapless Dirac system is conserved even in the presence of Coulomb interaction. This is equivalent to saying that perfect backscattering is forbidden in Dirac systems, and as a consequence the interference between the forward and backward propagating disturbance, which leads to Friedel oscillation in 1d,  does not occur in $1$d massless Dirac systems. 
Another interesting and well known effect related to the conservation of chirality in $1$d and lack of perfect backscattering in  massless Dirac systems is the phenomena of perfect Klein tunneling  [\onlinecite{KT, klein1}]. 

\subsection{Plasmons}
The polarization function in the dynamic long wavelength limit $q \to 0$ first and then $\omega \to 0$, {\it i.e.,} with ${\tilde \omega} > q$ fixed, which is useful to obtain the long wavelength plasmon dispersion, can be obtained from Eqs.~\eqref{chi1Dr2}-\eqref{chi1Dr1}, and it is given by 
\be \label{Pl1d}
\Pi^{(1d)}(q \to 0,{\tilde \omega})= \frac{g}{\pi\hbar v_{\rm F}}\frac{k_{\rm F}}{\sqrt{k_{\rm F}^2+{\tilde \Delta}^2}}\frac{q^2}{{\tilde \omega}^2} + {\cal O}(q^4/\omega^4)~.
\ee
 As a consequence the long wavelength limit of the plasmon dispersion is given by 
\be
\omega_{\rm pl} ^{(1d)} =  \sqrt{ \frac{2 g e^2 }{ \pi \hbar v_{\rm F}\kappa }} \hbar v_{\rm F} q \sqrt{ \frac{\hbar v_{\rm F} k_{\rm F} K_0 (qa)}{\mu} }
+ {\cal O}(q^3)~, 
 \label{wm1}
\ee
which is consistent with the results of Ref.~ [\onlinecite{rashi}]. The plasmon dispersion is displayed in the backdrop of the p-h continuum {  ({\it i.e.}, where the imaginary contribution of polarization function is non zero) in Fig.~\ref{fig2}(b). At $T=0$, the plasmon mode remains undamped for a wide range of $q$ and $\omega$, and then it becomes damped as it enters the p-h region (blue shaded region of Fig.~\ref{fig2}(b)) and decays by creating 
single p-h excitations}.
\subsection{Massless 1d electrons (${\tilde \Delta} \to 0$)}
In the ${\tilde \Delta} \to 0$ limit in 1d, the total polarization function reduces to 
\be \label{Eq:pi1d0}
\Pi^{(1d)}~(q,{\tilde \omega},{\tilde \Delta}=0)=\frac{g}{\pi {\hbar v_{\rm F}}}\frac{q^2}{({\tilde \omega} + i \eta)^2-q^2}~,
\ee
where the whole contribution arises from the intrinsic part. This can also be deduced from Eqs.~\eqref{chi1Dr2}-\eqref{chi1Dr1}, where all the terms proportional to $\Delta$ vanish, and only the $h_1 (q, \tilde \omega)$ term in Eq.~\eqref{chi1Dr2} contributes. Note that in this case, the imaginary part of the polarization function is non-zero only around the line ${\tilde \omega} = q$, {\it i.e.},  
\be
\Im m~\Pi^{(1d)}~(q,{\tilde \omega},{\tilde \Delta}=0)= \frac{- g}{{\hbar v_{\rm F}}}q^2 {\delta}({\tilde \omega} -q),
\ee
as shown in Fig.~\ref{fig2}(a)
This has two remarkable consequences, 
(1) since there is no non-analyticity in the static response function [see  Fig.~\ref{fig2}(e)], there are no Friedel oscillations in the density due to a charged impurity as mentioned earlier in Sec.~\ref{static1D}, and (2) the phase space for p-h excitations is drastically reduced here since only chirality conserving p-h excitations are allowed, as opposed to one dimensional systems with massive Dirac dispersion or parabolic dispersion.  
Further in the static limit $\omega \to 0$, the right hand side of Eq.~\eqref{Eq:pi1d0}, reduces to $-g/(\pi \hbar v_F)$.

For the case of massless Dirac fermions in $1$d, we can easily obtain the exact plasmon dispersion for a given form of the electrostatic potential $V_q$ to be
\be \label{pl_gapless}
\omega_{\rm pl}^{(1d)}(\Delta \to 0 ) = \hbar v_{\rm F} q \left( 1 + \frac{g}{\pi {\hbar v_{\rm F}}} V_q \right)^{1/2}~.
\ee
Note that since the p-h continuum is almost non-existent in 1d gapless Dirac systems [except along the $\omega = v_{\rm F} q$ line as shown in Fig.~\ref{fig2}(a)], the plasmon dispersion will be practically undamped with a very high quality factor.

Before proceeding to the case of 2d massive Dirac Fermions, we note that RPA or the treatment of the electron-electron interaction at the mean field level retaining only Hartree contribution, works much better in higher dimensions as opposed to 1d, where the Luttinger Liquid theory [\onlinecite{Giamarchi}] works well to describe the interacting itinerant electrons. In fact we already discussed in Sec.~III A that RPA fails to describe the Friedel oscillations (generally a large wave-vector phenomena involving $q = 2 k_{\rm F}$) for  1d interacting electrons. However RPA still gives a good estimate of the plasmon mode in 1d systems, particularly in the low wave-vector $q \to 0$ limit (homogeneous system response), as evidenced by the good match of the RPA results for the plasmon dispersion in the 1d parabolic case, with the corresponding experimental results in Ref.~[\onlinecite{Goni}]. More importantly, Luttinger liquid theory is based on the linearized spectrum of the p-h continuum around the Fermi point, and consequently the plasmons within the Luttinger Liquid will always be damped. However, within RPA, the plasmons can also lie outside the regime of 
p-h continuum, and will be completely undamped in that regime [\onlinecite{1D1, 1D2,SDS2,rashi}]. 
%[Phys. Rev. B 32, 1401(R) (1986)], [Phys. Rev. B.43, 11768 (1991)], [Phys. Rev. Lett. 102, 206412 (2009)] and [Phys. Rev. B 91, 205426 (2015)].

\section{Polarization function of 2d massive Dirac systems}\label{polfunc2d}
Having studied the 1d massive Dirac case, we now focus on the 2d case of gapped Dirac fermions. The polarization function for this case has already been derived in Ref.~[\onlinecite{Pyatkovskiy}], and we reproduce the results here for completeness. Let us first consider the intrinsic part, $\Pi^0 (q,{\tilde \omega})$, whose real and imaginary parts can be calculated following Ref.~[\onlinecite{Pyatkovskiy}], and combined to obtain, 
\bea \label{pol2dint}
\Pi_{0}^{(2d)}(q,{\tilde \omega}) &=& \frac{-g q^2}{8\pi \hbar v_{\rm F} (q^2-{\tilde \omega}^2)}\bigg(2{\tilde \Delta}+\frac{q^2-{\tilde \omega}^2-4{\tilde \Delta}^2}{\sqrt{q^2-{\tilde \omega}^2}}\nonumber\\
 & \times& \text{arcsin}\sqrt{\frac{q^2-{\tilde \omega}^2}{q^2-{\tilde \omega}^2+4{\tilde \Delta}^2}}\bigg)~. 
\eea

For evaluating the extrinsic part of the polarization function (for ${ \mu}>{ \Delta}$)  we split the $\chi_\mu^{\pm}$ integrals of Eq.~\eqref{split}  into real and imaginary parts using the Dirac identity. Thereafter doing some algebraic simplifications and by introducing the following notation, 
\bea \label{functions}
f(q, {\tilde \omega})&=& -\frac{g q^2}{16\pi\sqrt{|q^2-{\tilde \omega}^2|}}~,\nonumber\\
G_<(x)&=& x\sqrt{x_0^2-x^2}-(2-x_0^2)\text{arccos}(x/x_0)~, \nonumber\\
G_>(x)&=& x\sqrt{x^2-x_0^2}-(2-x_0^2)\text{arccosh}(x/x_0)~, \nonumber\\
G_0(x)&=& x\sqrt{x^2-x_0^2}-(2-x_0^2)\text{arcsinh}(x/\sqrt{-x_0^2})~,\nonumber\\
\eea
where $x_0$ is defined in Eq.~\eqref{x0},  we find the real part of the extrinsic polarization function to be 
 \bea \label{2D_real}
& & {\Re e~\Pi_{1}^{(2d)}(q,{\tilde \omega})} =-\frac{g{(\tilde \mu-\tilde \Delta)}}{2\pi \hbar v_{\rm F}}-\frac{f(q, {\tilde \omega})}{\hbar v_{\rm F}}\hspace{2cm}\nonumber\\ & & \times
\begin{cases}
-G_<\left(\frac{2{\tilde \Delta}-{\tilde \omega}}{q}\right)-G_<\left(\frac{2{\tilde \Delta}+{\tilde \omega}}{q}\right),  &\text{1A}\\
G_<\left(\frac{2{\tilde \mu}-{\tilde \omega}}{q}\right)-G_<\left(\frac{2{\tilde \Delta}-{\tilde \omega}}{q}\right)\\
-G_<\left(\frac{2{\tilde \Delta}+{\tilde \omega}}{q}\right), &\text{2A}\\
G_<\left(\frac{2{\tilde \mu}+{\tilde \omega}}{q}\right)+G_<\left(\frac{2{\tilde \mu}-{\tilde \omega}}{q}\right)\\
-G_<\left(\frac{2{\tilde \Delta}+{\tilde \omega}}{q}\right)-G_<\left(\frac{2{\tilde \Delta}-{\tilde \omega}}{q}\right), &\text{3A}\\
G_<\left(\frac{2{\tilde \mu}-{\tilde \omega}}{q}\right)-G_<\left(\frac{2{\tilde \mu}+{\tilde \omega}}{q}\right)\\
- G_<\left(\frac{2{\tilde \Delta}+{\tilde \omega}}{q}\right)-G_<\left(\frac{2{\tilde \Delta}-{\tilde \omega}}{q}\right), &\text{4A}\\
G_>\left(\frac{2{\tilde \mu}+{\tilde \omega}}{q}\right)-G_>\left(\frac{2{\tilde \mu}-{\tilde \omega}}{q}\right)\\
-G_>\left(\frac{2{\tilde \Delta}+{\tilde \omega}}{q}\right)+G_>\left(\frac{{\tilde \omega}-2{\tilde \Delta}}{q}\right), &\text{1B}\\
G_>\left(\frac{2{\tilde \mu}+{\tilde \omega}}{q}\right)-G_>\left(\frac{2{\tilde \Delta}+{\tilde \omega}}{q}\right)\\
-G_>\left(\frac{2{\tilde \Delta}-{\tilde \omega}}{q}\right), &\text{2B}\\
G_>\left(\frac{2{\tilde \mu}+{\tilde \omega}}{q}\right) - G_>\left(\frac{{\tilde \omega}-2{\tilde \mu}}{q}\right)\\
-G_>\left(\frac{2{\tilde \Delta}+{\tilde \omega}}{q}\right)+ G_>\left(\frac{{\tilde \omega}-2{\tilde \Delta}}{q}\right), &\text{3B}\\
G_>\left(\frac{2{\tilde \mu}+{\tilde \omega}}{q}\right) + G_>\left(\frac{{\tilde \omega}-2{\tilde \mu}}{q}\right)\\
+G_>\left(\frac{2{\tilde \Delta}+{\tilde \omega}}{q}\right) -G_>\left(\frac{{\tilde \omega}-2{\tilde \Delta}}{q}\right), &\text{4B}\\
G_0\left(\frac{2{\tilde \mu}+{\tilde \omega}}{q}\right)-G_0\left(\frac{2{\tilde \mu}-{\tilde \omega}}{q}\right)\\
-G_0\left(\frac{2{\tilde \Delta}+{\tilde \omega}}{q}\right)+G_0\left(\frac{2{\tilde \Delta}-{\tilde \omega}}{q}\right), &\text{5B}~.
\label{pol2dextreal}\end{cases}
\eea
where the various regions in the ${\tilde \omega}-q$ plane have been defined in Eq.~\eqref{eq:regions} and marked in Fig.~\ref{fig1}(b). 
A similar calculation for the imaginary part of the extrinsic polarization function leads to
\bea \label{2D_imag}
& & \Im m ~\Pi^{(2d)}_1(q, {\tilde \omega},)= \frac{f(q, {\tilde \omega})}{ \hbar v_{\rm F}} \\ 
& & \nn
\times
\begin{cases}
G_>\left(\frac{2{\tilde \mu}+{\tilde \omega}}{q}\right)-G_>\left(\frac{2{\tilde \mu}-{\tilde \omega}}{q}\right),  &\text{1A}\\
G_>\left(\frac{2{\tilde \mu}+{\tilde \omega}}{q}\right), &\text{2A}\\
-\pi(2-x_0^2), &\text{1B}\\
-G_<\left(\frac{2{\tilde \mu}-{\tilde \omega}}{q}\right)-\pi(2-x_0^2), &\text{2B}\\
0 , &\text{3A, 3B, 4A, 4B, 5B} \\
%0 , & \text{4A, 4B, 5B}~.
\label{pol2dextimag}\end{cases}
\eea 
Note that the real and imaginary part of the intrinsic polarization function [see Eq.~\eqref{pol2dint}] can also be split  region-wise as
\bea \label{2D_intreal}
& & \Re e~\Pi^{(2d)}_0(q,{\tilde \omega})=\frac{-g\Delta}{2\pi \hbar v_{\rm F}} + \frac{f(q,{\tilde \omega})}{\hbar v_{\rm F}} \\ 
& & \times \nn
\begin{cases}
-G_>\left(\frac{2\tilde{\Delta}+\tilde{\omega}}{q}\right)-G_>\left(\frac{2{\tilde \Delta}-{\tilde \omega}}{q}\right)~, &\text{1B,2B,3B,4B}\\
-G_0\left(\frac{2{\tilde \Delta} + {\tilde \omega}}{q}\right)+G_0\left(\frac{2{\tilde \Delta} - {\tilde \omega}}{q}\right)~,&\text{5B}\\
-G_<\left(\frac{2{\tilde \Delta} - {\tilde \omega}}{q}\right)-G_<\left(\frac{2{\tilde \Delta} + {\tilde \omega}}{q}\right)~,&\text{1A,2A,3A,4A}\\
\end{cases}
\eea
where the $G,~G_<$, and $G_>$ functions are defined in Eq.~\eqref{functions}, and 
\bea \label{2d_intimag}
\Im m ~\Pi^{(2d)}_{0}~(q,{\tilde \omega}) &=& \frac{f(q,{\tilde \omega})}{\hbar v_{\rm F}} 
\\ & \times &
\begin{cases}
\pi (2-x_0^2)~,&\text{1B,2B,3B,4B} \\
0~,&\text{5B,1A,2A,3A,4A} \nn
\end{cases}
\eea
Adding Eq.~\eqref{2D_real} to Eq.~\eqref{2D_intreal}, and Eq.~\eqref{2D_imag} to Eq.~\eqref{2d_intimag} respectively, gives the 
total polarization function which is identical to that reported in Ref.~[\onlinecite{Pyatkovskiy}]. 
In the gapless limit ($\Delta \to 0$), the polarization function reduces to the well known polarization function of graphene [\onlinecite{Guinea_NJP2007, SDS1}]. 
We now use these analytical expressions to study the static limit, and calculate the plasmon dispersion and its long wavelength limit in subsequent subsections. 

\subsection{Static dielectric function and Friedel oscillations}
The intrinsic part of the Lindhard function for $\mu<\Delta$ in the static limit (${\tilde \omega}=0$) can be obtained from 
Eq.~\eqref{pol2dint}, and it is given by
\be \label{pol2dint2}
{\Pi_{0}^{(2d)}(q, 0)} = -\frac{g}{8\pi {\hbar v_{\rm F}}}\bigg(2{\tilde \Delta}+\frac{q^2-4{\tilde \Delta}^2}{q} \text{arcsin}\frac{q}{\sqrt{q^2+4{\tilde \Delta}^2}}\bigg)~. 
\ee
The static limit for the extrinsic polarization function ($\mu>\Delta$) can be obtained using Eqs. ~\eqref{pol2dextreal}-\eqref{pol2dextimag}, and added to Eq.~\eqref{pol2dint2} to obtain the total static polarization, 
\bea {\Pi^{(2d)}(q, 0)} &=& -\frac{g{\tilde \mu}}{2 \pi {\hbar v_{\rm F}}}\bigg[1-\theta(q-2k_{\rm F})\bigg(\frac{\sqrt{q^2-4k_{\rm F}^2}}{2q}\nonumber\\
&-& \frac{q^2-4{\tilde \Delta}^2}{4q{\tilde \mu}}~ \text{arctan}\frac{\sqrt{q^2-4k_{\rm F}^2}}{2{\tilde \mu}}\bigg)\bigg]\label{pol2dextstatic}~. 
\eea

For a localized charged impurity placed in a $2$d Dirac electron liquid, the oscillating screened potential far away from the impurity location,  is given by (see appendix \ref{FO}), 
%\be \label{eq:2dgapfriedel}
%\phi(r)\approx 4k_{\rm F}^2~A^{(2)} \frac{\sin(2k_{\rm F}r)}{(2k_{\rm F}r)^2} \frac{1}{2\sqrt{2}k_{\rm F}}\left(1-\frac{k_{\rm F}^2-\tilde{\Delta}^2}{{\tilde \mu}^2}\right)~,
%\ee
\be \label{eq:2dgapfriedel}
\phi(r)\approx A^{(2)} \frac{g{\tilde \mu}k_{\rm F}}{\hbar v_{\rm F}} \frac{\sin(2k_{\rm F}r)}{(2k_{\rm F}r)^2} \left(1-\frac{k_{\rm F}^2-\tilde{\Delta}^2}{{\tilde \mu}^2}\right)~,
\ee
where $A^{(2)}$ is a constant given by Eq.~\eqref{A}. The screened potential in the $2$d massive Dirac case, decays asymptotically as  $r^{-2}$, similar to the case of a parabolic $2$d electron gas, and unlike that in 2d gapless Dirac electron gas, in which the decay rate is proportional to $r^{-3}$ [\onlinecite{Guinea_NJP2007}]. 
This is also evident from Eq.~\eqref{eq:2dgapfriedel} in which the $r^{-2}$ dependence arises from the first order term in the expansion of 
$\arctan(\sqrt{q-2k_{\rm F}})$, and in which the right hand side vanishes for the gapless case {\it i.e.}, $\Delta \to 0$. Following the methodology of appendix \ref{FO}, it is easy to show that the main contribution in the gapless case arises from the  second order term in expansion of $\arctan(\sqrt{q-2k_{\rm F}})$ and in this case the screened potential in the asymptotic limit  is given by [\onlinecite{Pyatkovskiy,Schliemann3}]
\be
\phi(r) \approx   A^{(2)}_{\Delta = 0}~\frac{g{\tilde \mu}k_{\rm F}}{\hbar v_{\rm F}} \frac{\cos(2k_{\rm F}r)}{(2k_{\rm F}r)^3}~.
\ee
{ where $A^{(2)}_{\Delta = 0}$ is a constant given by Eq.~\eqref{A} with the substitution $\Delta=0$}. The different scaling of the screening potential is a consequence of the fact that in the gapless case the polarization function at $q=2k_{\rm F}$ has a discontinuity in second derivative, while in the gapped case the polarization function has a discontinuity in the  first derivative at $q=2k_{\rm F}$. 
%\iffalse
\begin{figure}[t] 
\begin{center} 
\includegraphics[width=0.95\linewidth]{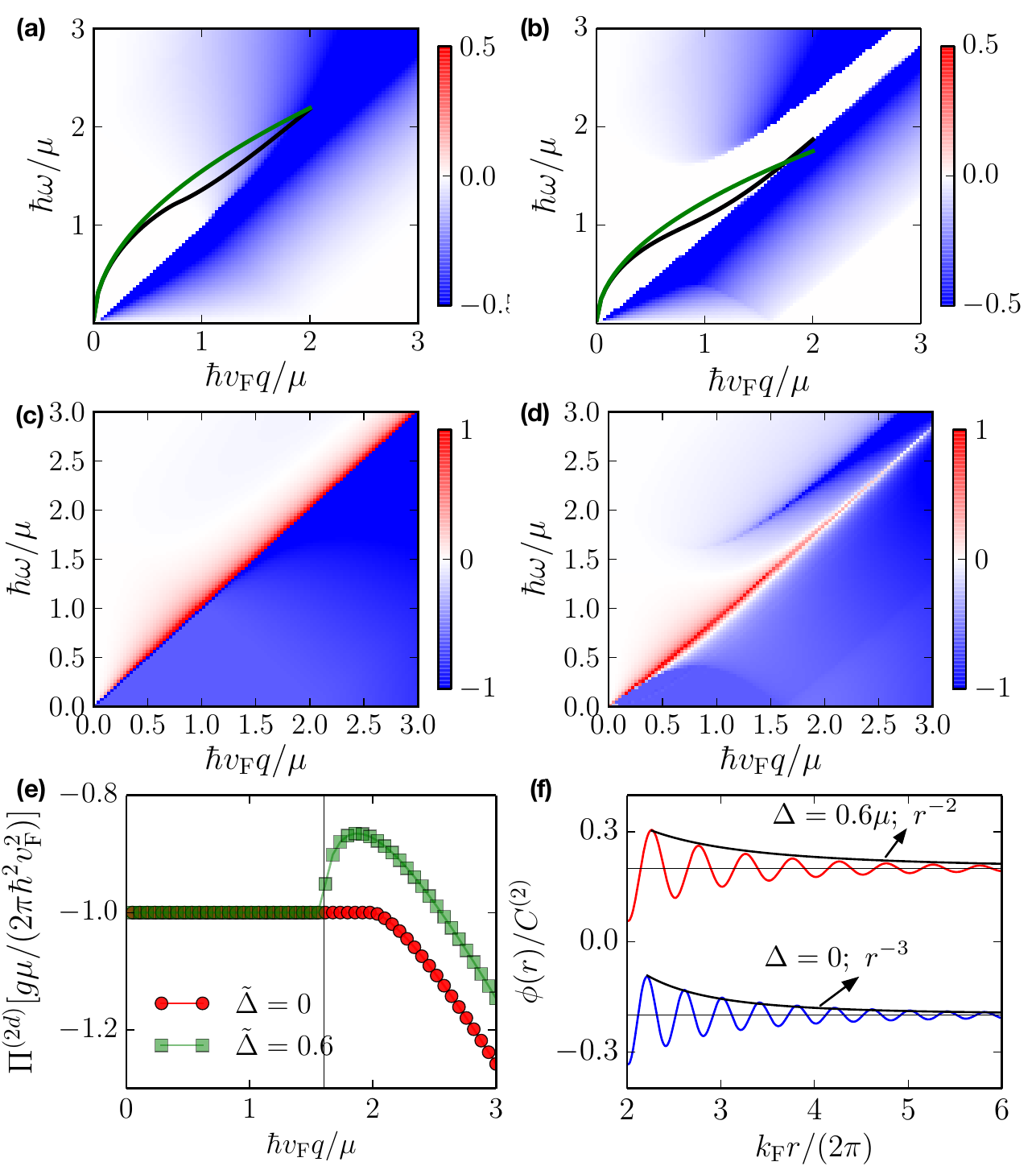} 
\end{center} 
\caption{(a)${\Im m \Pi(q,{\tilde \omega})}$ in the ${\tilde \omega}-q$ plane,  for a massless $2$d Dirac system along with the exact (black curve) and the long wavelength (green curve) plasmon dispersion,  
(b) ${\Im m \Pi(q,{\tilde \omega})}$ for a massive $2$d Dirac system along with the exact and the long wavelength plasmon dispersion.  (c) - (d)  ${\Re e \Pi(q,{\tilde \omega})}$ for $\Delta = 0$ and $\Delta = 0.6 \mu$, respectively. We have chosen $e^2 k_{\rm F}/(\hbar \kappa v_{\rm F} ) =  1.2/\pi$ for calculating the plasmon dispersion and in panels (a)-(d) the polarization function is in units of {$ \mu/{(\hbar^2 v_{\rm F}^2)}$}. % In panels (a)-(d) the polarization function is in units of $g/(4 \pi)$. }
(e) The static $\Pi(q,{\tilde \omega}=0)$ versus $q$, for the massless and massive Dirac system in $2$d.  
 (f) Asymptotic behaviour of the Friedel oscillations in the potential of a charged impurity in a massive (red curve, shifted up by 0.1 units) and massless (blue curve, shifted down by 0.1 units) $2$d Dirac system.    Here $C^{(2)} = A^{(2)} \times g \mu/(\pi \hbar^2 v_{\rm F}^2)$.
\label{fig3} } 
\end{figure}
%\fi

\subsection{Plasmons}
 To calculate the frequency of the collective density excitations {\it i.e.}, the plasmon dispersion, we solve Eq.~\eqref{eq:condition} numerically. 
 The numerically calculated plasmon dispersion is displayed in Fig.~\ref{fig3}b, via solid black line, on top of the imaginary part of the polarization function which depicts various regions of the p-h continuum. Note that the imaginary part of the polarization function is non-zero in  regions $1$A,$2$A, $2$B and $4$B of Fig.~\ref{fig1}(b), and hence these regions form the single particle excitation spectrum, as marked by shaded area in Fig. \ref{fig3}(c). 
In the dynamic long wavelength limit, $q \to 0$ first and then $\omega \to 0$, {\it i.e.,} with ${\tilde \omega} > q$ fixed, 
the polarization function is given by 
\be
\Pi^{(2d)}(q \to 0,{\tilde \omega})= \frac{g}{2\pi\hbar v_{\rm F}}\frac{k_{\rm F}^{2}}{\sqrt{k_{\rm F}^2+{\tilde \Delta}^2}}\frac{q^2}{{\tilde \omega}^2} + {\cal O}(q^4/\omega^4)~.
\ee 
Using the above expression for the polarization function, we find the long wavelength plasmon dispersion to be 
\be \label{2dlw}
\omega^{(2d)}_{\rm pl} (q \to 0)= \sqrt{\frac{g e^2 \mu q}{2 \kappa\hbar^2}}\left(1-\frac{\Delta^2}{\mu^2}\right)^{1/2}~. 
\ee 
Note that similar to the case for massless Dirac fermions, and parabolic dispersion systems in two dimensions, the plasmon dispersion in the long wavelength limit is $\propto q^{1/2}$. This is a consequence of the charge continuity equation, and the form of the bare Coulomb interaction in the momentum space in the long wavelength limit in 2d. 
However the density dependent pre-factor in all the three cases are different, and depend on the details of the electronic dispersion relation, doping etc. of the system considered. 
\subsection{Massless 2d electrons, $\Delta \to 0$}
In the massless limit ($\Delta \to 0$) for Dirac fermions in $2$d, $x_0$ becomes equal to $1$, hence regions $4$A, $4$B, and $5$B vanish [see Fig.~\ref{fig1}(a)], and the real and imaginary parts of the total polarization function can be combined  to obtain [\onlinecite{Pyatkovskiy}],
\be \label{eq:gr}
\frac{\Pi^{(2d)}(q,{\tilde \omega})}{N_0} = -\frac{{\tilde \mu}}{2k_{\rm F}} + F(q, {\tilde \omega}) \Big[ G_+(\nu_{+}) + G_-(\nu_{-})\Big]~,
\ee
where $N_0 = g k_{\rm F}/(\pi \hbar v_{\rm F})$ is the density of states at the Fermi surface of a $2$d gapless system, and we have defined the dimensionless variables as~$\nu_{\pm} = 2 v_{\rm F}/{q} \pm {{\tilde \omega}}/q~$.  Additionally in Eq.~\eqref{eq:gr}, we have used the following functions,   
\be 
F(q,{\tilde \omega}) = \frac{q^2}{16 k_{\rm F} \sqrt{q^2- {\tilde \omega}^2}}~,
\ee
and 
\be \label{eq:G}
G_{\pm}(z) = z \sqrt{1-z^2} \pm i \cosh^{-1}(z)~.
\ee

The polarization function in the static limit, $\omega \to 0$, is displayed in Fig.~\ref{fig3}(e), and it remains constant at $2 k_{\rm F}^2/\mu^2$ for $k < 2 k_{\rm F}$. The $q \to 0$ limit of the plasmon dispersion is  easily obtained from Eq.~\eqref{2dlw}, by setting $\Delta \to 0$. 
The exact plasmon dispersion calculated numerically, and the long wavelength results, are shown in Fig.~\ref{fig3}(a), in black and green solid lines,  respectively.  { Clearly the exact plasmon dispersion enters the p-h continuum at a larger $q$ (and lower $\omega$), as compared to the long wavelength result. In Fig.~ \ref{fig3}(b), the exact plasmon dispersion becomes almost parallel to the p-h boundary and enters the p-h continuum at a much larger $q$ and $\omega$ value, as opposed to the long wavelength result.}

\section{Polarization function of 3d massive Dirac systems}\label{polfunc3d}

In this section, we calculate the polarization function for $3$d massive Dirac materials. Similar to $1$d and $2$d calculations, we evaluate the intrinsic and extrinsic parts of the polarization function separately.  
The general form of intrinsic $\Pi_0^{(3d)}(q,\tilde{\omega})$ and extrinsic $\Pi_1^{(3d)}(q,\tilde{\omega})$ polarization function following from Eq. (\ref{split}) and after integration over azimuthal angle $\phi$ 
and upon simplifying the $\theta$ integration is given by
% \bea 
% \Pi_{0}^{(3d)}(q,{\tilde \omega}) &=&  \frac{ g}{16\pi^2 \hbar v_{\rm F}} \int_{0}^{\Lambda} \frac{k~ dk}{q ~\tilde{E}_{\textbf{k}}} \int_{l_1}^{l_2} dk'  \Bigg[\left(-3\tilde{E}_{\bf k} + k'\right) \nn \\
% & +& \frac{(2\tilde{E}_{\bf k} + \tilde{\omega})^2 - q^2}{\tilde{E}_{\bf k} + k' + {\tilde \omega}} \Bigg] + [{\tilde \omega} \to -{\tilde \omega} ]\label{eq:Pi_int_3d}\\
% \Pi_{1}^{(3d)}(q,{\tilde \omega}) &=& \frac{ g }{16\pi^2 \hbar v_{\rm F}} \int_{0}^{k_{\rm F}} \frac{k~ dk}{q ~\tilde{E}_{\textbf{k}}} \int_{l_1}^{l_2} dk' \Bigg[ -2k'\nn \\
% &+&\frac{2k'[(2\tilde{E}_{\bf k} + {\tilde \omega})^2 - q^2]}{(\tilde{E}_{\bf k}+ \tilde{\omega})^2 -k'^2}\Bigg] + [\tilde{\omega} \to -{\tilde \omega}]~, \nn \\ \label{eq:Pi_ext_3d}
% \eea
{ 
\bea
\Pi_{0}^{(3d)}(q,{\tilde \omega}) &=& I(q,{\tilde \omega})+I(q,-{\tilde \omega})\label{eq:Pi_int_3d}~,~~~{\rm and}\\
\Pi_{1}^{(3d)}(q,{\tilde \omega}) &=& J(q,{\tilde \omega})+J(q,-{\tilde \omega})\label{eq:Pi_ext_3d}~,
\eea
where
\bea
I(q,{\tilde \omega}) &=& \frac{ g}{16\pi^2 \hbar v_{\rm F}} \int_{0}^{\Lambda} \frac{k~ dk}{q ~\tilde{E}_{\textbf{k}}} \int_{l_1}^{l_2} dk'  \Bigg[\left(-3\tilde{E}_{\bf k} + k'\right) \nn \\
& +& \frac{(2\tilde{E}_{\bf k} + \tilde{\omega})^2 - q^2}{\tilde{E}_{\bf k} + k' + {\tilde \omega}} \Bigg] , ~~~{\rm and} \label{II} \\
J(q,{\tilde \omega}) &=& \frac{ g }{16\pi^2 \hbar v_{\rm F}} \int_{0}^{k_{\rm F}} \frac{k~ dk}{q ~\tilde{E}_{\textbf{k}}} \int_{l_1}^{l_2} dk' \Bigg[ -2k'\nn \\
 &+&\frac{2k'[(2\tilde{E}_{\bf k} + {\tilde \omega})^2 - q^2]}{(\tilde{E}_{\bf k}+ \tilde{\omega})^2 -k'^2}\Bigg]~. \label{JJ}
\eea 
In Eqs.~ \eqref{II}-\eqref{JJ}, we have} $\tilde {\omega} \to \tilde{\omega} + i \eta$, $l_1=\sqrt{|k-q|^2+{\tilde \Delta}^{2}}$, $l_2=\sqrt{(k+q)^2+{\tilde \Delta}^{2}}$, $\Lambda$ is the large momentum cut off and $k'$ originally was $\tilde{E}_{\textbf{k+q}}$. 
Calculating the imaginary part of the intrinsic polarization function using the Dirac identity, we obtain
\be \label{im03d} 
\Im m~\Pi_{0}^{(3d)} = \frac{ gq^2}{48 \pi \hbar v_{\rm F}}x_0 \left(x_0^2-3\right) \theta ({\tilde \omega}^2-q^2-4{\tilde \Delta}^2)  ~, \\ 
\ee
where $x_0$ is specified by Eq.~\eqref{x0}.
Similarly the imaginary component of the extrinsic part of the polarization function is given by 
{\be \label{im13d}
\Im m~\Pi_1^{(3d)}=\frac{-g }{8\pi \hbar v_{\rm F}}
\begin{cases}
G(q,\tilde{\omega})-G(q,-\tilde{\omega}),~~~~ \text{1A} \\
G(q,\tilde{\omega})-\frac{q^2}{12}(2+x_0^3-3x_0), ~~~~ \text{2A} \\
\frac{q^2}{6}(x_0^3-3x_0), ~~~~ \text{1B} \\
-G(-q,-\tilde{\omega})+\frac{q^2}{12}(2+x_0^3-3x_0), ~~~~ \text{2B} \\
0,~~~~ \text{3A}, \text{3B}, {\text{4A}, \text{4B},  \text{5B}}~,\\
\end{cases}~
\ee}
%\be \label{im13d}
%\Im m~\Pi_1^{(3d)}={ \frac{-g }{16\pi \hbar v_{\rm F}}}
%\begin{cases}
%2G(q,-\tilde{\omega})+2G(q,\tilde{\omega})-\frac{q^2}{3}(2+x_0^3-3x_0),~~~~ \text{1A} \\
%2G(q,\tilde{\omega})-\frac{q^2}{6}(2+x_0^3-3x_0), ~~~~ \text{2A} \\
%\frac{q^2x_0}{3}(x_0^2-3), ~~~~ \text{1B} \\
%-2G(-q,-\tilde{\omega})+\frac{q^2}{6}(2+x_0^3-3x_0), ~~~~ \text{2B} \\
%0,~~~~ \text{3A}, \text{3B}, {\text{4A}, \text{4B},  \text{5B}}~,\\
%\end{cases}~
%\ee
where we have defined the function 
\be \label{eq:G}
G(q,\tilde{\omega})=\frac{1}{12q}[(2{\tilde \mu}+{\tilde \omega})^3-3q^2(2{\tilde \mu}+{\tilde \omega})+2q^3]~.
\ee
As a consistency check we note that in the $\Delta \to 0$ limit, Eq. (\ref{im03d}) and  Eq.~\eqref{im13d}, reproduce the results reported earlier for the massless case --- see Eq.~(A16) of 
Ref.~[\onlinecite{xiao_arxiv_2014}].

For the real part of the polarization function, we integrate the principal value of the integrand over $k'$ in Eqs.~(\ref{eq:Pi_int_3d})-(\ref{eq:Pi_ext_3d}) to obtain the following, 
\begin{widetext}
\bea \label{eq:3d_kint}
\Re e~\Pi_0^{(3d)}(q,{\tilde \omega}) &=& \frac{g}{16\pi^2q \hbar v_{\rm F}} \int_{0}^{\Lambda} \frac{k dk}{{\tilde E}_{\textbf{k}}}\left[ 2kq+3{\tilde E}_{\textbf{k}}\sqrt{(k-q)^2+\tilde{\Delta}^2}-3{\tilde E}_{\textbf{k}}\sqrt{(k+q)^2+\tilde{\Delta}^2}+\{(2{\tilde E}_{\textbf{k}}+{\tilde \omega})^2-q^2\} \right.\nn \\
&\times& \left. \log\left|\frac{{\tilde E}_{\textbf{k}}+\sqrt{(k+q)^2+{\tilde \Delta}^2}+{\tilde \omega}}{{\tilde E}_{\textbf{k}}+\sqrt{(k-q)^2+{\tilde \Delta}^2}+{\tilde \omega}}\right|\right] + {\tilde \omega} \to -{\tilde \omega}~,  \\ \label{eq:3d_kint2}
\Re e~\Pi_1^{(3d)}(q,{\tilde \omega}) &=& \frac{g}{16\pi^2q \hbar v_{\rm F}} \int_{0}^{k_{\rm F}} \frac{k dk}{{\tilde E}_{\textbf{k}}}\left[-4kq+\{(2\tilde{E}_\textbf{k}+\omega)^2-q^2\} \log\left|\frac{(\tilde{E}_\textbf{k}+\tilde{\omega})^2-\{(k-q)^2+\tilde{\Delta}^2\}}{(\tilde{E}_\textbf{k}+\tilde{\omega})^2-\{(k+q)^2+\tilde{\Delta}^2\}}\right|\right] + {\tilde \omega} \to -{\tilde \omega}~. \nonumber \\
\eea
\end{widetext}
While the integrals in Eqs.~\eqref{eq:3d_kint}-\eqref{eq:3d_kint2} can be done analytically, the resulting expressions are very cumbersome, and do not offer any useful insight. Additionally, we find it easier to do these integrals numerically and use the numerical results to obtain the plasmon dispersion and other limiting cases thereafter.
We note that in the limiting case of massless Dirac systems {\it i.e.}, $\Delta=0$ implying $x_0=1$, 
Eqs.~\eqref{eq:3d_kint}-\eqref{eq:3d_kint2} reproduce the known result for the real part of the polarization function for a gapless $3$d Dirac system [\onlinecite{xiao_arxiv_2014}].

\subsection{Static dielectric function and Friedel oscillations}
In the limiting case of $\omega=0$ at  finite $q$, the imaginary part of the static polarization vanishes, and it reduces to 
\begin{widetext}
\bea \label{eq:3dstat0}
\Pi_0^{(3d)}~(q,{\tilde\omega}=0)&=&\frac{g}{48\pi^2 \hbar v_{\rm F}q}\Big[8q{\tilde \Delta}^2+4q^3 \log \frac{{\tilde \Delta}}{2\Lambda}+\frac{2(q^4+2q^2{\tilde \Delta}^2-8{\tilde \Delta}^4)}{\sqrt{q^2+4{\tilde \Delta}^2}}\log \frac{q+\sqrt{q^2+4{\tilde \Delta}^2}}{-q + \sqrt{q^2+4{\tilde \Delta}^2}}\Big]~,
\eea
\bea\label{eq:3dstat1}
\Pi_1^{(3d)}~(q,{\tilde \omega}=0)&=&\frac{g }{24\pi^2 \hbar v_{\rm F} q}\Bigg[-8k_{\rm F}q\tilde{\mu}+\left(\tilde{\mu}(4k_{\rm F}^2-3q^2+4{\tilde \Delta}^2)+\frac{q^4+2q^2{\tilde \Delta}^2-8{\tilde \Delta}^4}{\sqrt{q^2+4{\tilde \Delta}^2}}\right)\log\left|\frac{q-2k_{\rm F}}{q+2k_{\rm F}}\right|\nn \\
&+& 2q^3\log\frac{k_{\rm F}+\tilde{\mu}}{{\tilde \Delta}}+\frac{q^4+2q^2\tilde{\Delta}^2-8\tilde{\Delta}^4}{\sqrt{q^2+4\tilde{\Delta}^2}}\log\frac{-k_{\rm F}q+2\tilde{\Delta}^2+\tilde{\mu}\sqrt{q^2+4\tilde{\Delta}^2}}{k_{\rm F}q+2\tilde{\Delta}^2+\tilde{\mu}\sqrt{q^2+4\tilde{\Delta}^2}}\Bigg]~.
\eea
\end{widetext}

The oscillating behaviour (Friedel Oscillations) of the screened potential in $3$d gapped Dirac system, obtained numerically from Eqs.~\eqref{eq:3dstat0}-\eqref{eq:3dstat1} and ~\eqref{eq:potential2}, is displayed in Fig.~\ref{fig4}(f) (red curve) and the $r^{-3}$ decay is evident. The $r^{-3}$ decay can also be inferred 
from the fact that  the static extrinsic polarization function in Eq.~(\ref{eq:3dstat1}), has a second derivative discontinuity at $q=2k_{\rm F}$,  from which we can estimate the asymptotic form of screened potential  to be 
\be 
\phi(r) \propto \frac{\sin(2k_{\rm F}r)}{r^3}~.
\ee 
In gapless $3$d system the screened potential can be calculated analytically using Eq.~\eqref{eq:potential2}, and it is given by 
\be \label{eq:3dFgapless}
\phi(r)\approx A^{(3)}\frac{2gk_{\rm F}^3}{\hbar v_{\rm F}}\frac{\sin(2k_{\rm F}r)}{(2k_{\rm F}r)^4}
\ee 
{ where $A^{(3)}$ is a constant given by Eq.~\eqref{A}}. Note that Eq.~\eqref{eq:3dFgapless} is a reproduction of  Eq.~(19) of Ref.[\onlinecite{zhang_ijmp_2013}].
The faster $r^{-4}$ decay in the Friedel oscillation of gapless 3d Dirac systems can be traced back to the fact that the static polarization function at $q=2k_{\rm F}$ has a discontinuity in third
derivative as compared to second derivative discontinuity at $q=2k_{\rm F}$ for gapped static response function.

\subsection{Plasmons}
\begin{figure}[t] 
\begin{center} 
\includegraphics[width=0.95\linewidth]{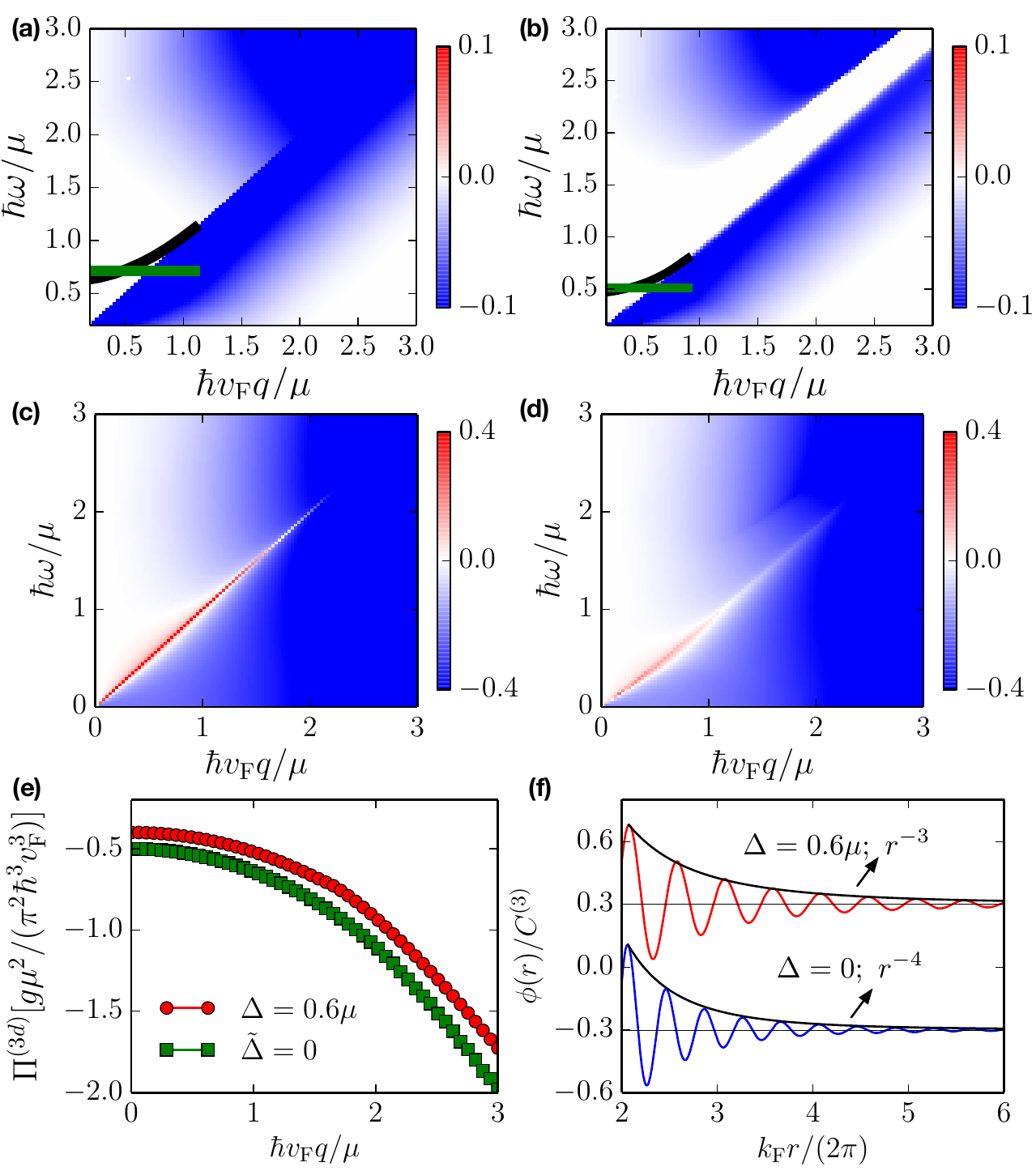} 
\end{center} 
\caption{(a) ${\Im m \Pi(q,{\tilde \omega})}$ in the ${\tilde \omega}-q$ plane,  for massless $3$d Dirac system along with the numerically exact (black curve) and the long wavelength (green curve) plasmon dispersion,  
(b) ${\Im m \Pi(q,{\tilde \omega})}$ for a massive $3$d Dirac system along with the exact and the long wavelength plasmon dispersion  (c) - (d)  ${\Re e \Pi(q,{\tilde \omega})}$ for $\Delta = 0$ and $\Delta = 0.6 \mu$, respectively. 
We have chosen $e^2 k_{\rm F}^2/(\hbar \kappa v_{\rm F} ) =  0.6\pi$ for calculating the plasmon dispersion. In panels (a)-(d) the polarization function is in units of $ \mu^2/{\hbar^3 v_{\rm F}^3}$. 
(e) The static $\Pi(q,{\tilde \omega}=0)$ versus $q$, for the massless and massive $2$d Dirac system.  
 (f) Asymptotic behaviour of the Friedel oscillations in the potential of a charged impurity in a massive (red curve, shifted up by 0.05 units) and massless (blue curve, shifted down by 0.05 units) $3$d Dirac system.  Here $C^{(3)} = A^{(3)} \times g \mu^2/(\pi^2 \hbar^3 v_{\rm F}^3)$. 
\label{fig4} } 
\end{figure}

The exact plasmon dispersion for the $3$d massive Dirac system is obtained by numerically solving 
Eq.~\eqref{eq:condition} and  is displayed in Fig.~\ref{fig4}(b), over the background of the $\Im m \Pi(q,\omega)$, which indicates the p-h continuum.  
The polarization function in the dynamic long wavelength limit $q \to 0$ first and then $\omega \to 0$, {\it i.e.,} with ${\tilde \omega} > q$ fixed,  is given by 
\be
\Pi^{(3d)}(q \to 0,{\tilde \omega})= \frac{g}{6\pi^2\hbar v_{\rm F}}\frac{k_{\rm F}^{3}}{\sqrt{k_{\rm F}^2+{\tilde \Delta}^2}}\frac{q^2}{{\tilde \omega}^2} + {\cal O}(q^4/\omega^4)~.
\ee
Using this expression for polarization function at small energies and momenta, we find the plasmon dispersion at long wavelength to be
\be
 \omega^{(3d)}_{\rm pl} = \sqrt{\frac{2ge^2\mu^2} {3 \pi\kappa \hbar^3 v_{\rm F}}}\left(1-\frac{\Delta^2}{\mu^2}\right)^{3/4}\label{3dlw}~, 
 \ee
 independent of $q$. This is consistent with the case of systems with parabolic dispersion in $3$d, and massless Dirac fermions in $3$d, as expected from the charge continuity equation and the form of the bare Coulomb interaction in momentum space in $3$d.  Note however, that at finite $q$, the plasmon dispersion develops a dependence on $q$, as evident from the exact numerical solution, displayed in Fig.~\ref{fig4}(a)-(b). {  Note that as opposed to the long wavelength result, in both panels (a) and (b) of Fig.~\ref{fig4}, the exact plasmon dispersion enters the p-h continuum and gets damped at a much larger $q$ and $\omega$ values. }
%
%\iffalse
%
%\fi
\subsection{Massless 3d electrons, $\Delta \to 0$}

In the massless limit for $3$d polarization function, the real and imaginary parts of the intrinsic polarization can be obtained as a limiting case of Eqs.~\eqref{im03d}  and \eqref{eq:3d_kint}, and are given by
\bea
\Re e~\Pi_{0}^{(3d)}(q,{\tilde \omega}) &=& -\frac{gq^2}{24\pi^2 \hbar v_{\rm F}} \log\bigg|\frac{4\Lambda^2}{q^2-{\tilde \omega}^2}\bigg| \label{3drepi0}~,\\
\Im m~\Pi_{0}^{(3d)}(q,{\tilde \omega}) &=& -\frac{gq^2\theta({\tilde \omega}-q)}{24\pi\hbar v_{\rm F}}\label{3dimpi0}~.
\eea
The real  part of the extrinsic polarization function is given by 
\bea
& &\Re e~\Pi_{1}^{(3d)} = \frac{gq^2}{8\pi^2\hbar v_{\rm F}}\bigg[-\frac{8{\tilde \mu}^2}{3q^2}+\frac{G(q,{\tilde \omega})H(q,{\tilde \omega})}{q^2} \nn\\
& &+\frac{G(-q,{\tilde \omega})H(-q,{\tilde \omega})}{q^2} +\frac{G(q,-{\tilde \omega})H(q,-{\tilde \omega})}{q^2}\nn\\
& &+ \frac{G(-q,-{\tilde \omega})H(-q,-{\tilde \omega})}{q^2}\bigg]\label{3drepi1}~,
\eea
where $G$ is specified by Eq.~\eqref{eq:G}, and $H$  is defined as  
\be 
H(q,{\tilde \omega})= \text{log}~\bigg|\frac{2{\tilde \mu}+{\tilde \omega}-q}{q-{\tilde \omega}}\bigg|~.
\ee
Similarly the imaginary part of the extrinsic polarization function can be obtained and it is given by 
{\be \label{im13d}
\Im m~\Pi_1^{(3d)}=\frac{-g }{8\pi \hbar v_{\rm F}}
\begin{cases}
G(q,\tilde{\omega})-G(q,-\tilde{\omega}),~~~~ \text{1A} \\
G(q,\tilde{\omega}), ~~~~ \text{2A} \\
-\frac{q^2}{3}, ~~~~ \text{1B} \\
-G(-q,-\tilde{\omega}), ~~~~ \text{2B} \\
0,~~~~ \text{3A}, \text{3B}, {\text{4A}, \text{4B},  \text{5B}}~,\\
\end{cases}~
\ee}
%{\bea
%& &\Im m~\Pi_{1}^{(3d)} = -\frac{gq^2}{8\pi^2 \hbar v_{\rm F}}\bigg[\theta(q-{\tilde \omega})\big[\frac{\pi G(q,{\tilde \omega})}{q^2}\theta(2{\tilde \mu}+{\tilde \omega}-q)\nn\\
%& &- \frac{\pi G(q,-{\tilde \omega})}{q^2}\theta(2{\tilde \mu}-{\tilde \omega}-q) \big]+\theta({\tilde \omega}-q)\nn\\
%& &\times \big[-\frac{\pi}{3}\theta(2{\tilde \mu}-{\tilde \omega}-q)- \frac{\pi G(-q,-{\tilde \omega})}{q^2}\theta(q+{\tilde \omega}-2{\tilde \mu}) \nn\\
%& &\times \theta(2{\tilde \mu}+q-{\tilde \omega})\big]\bigg] \label{3dimpi1}~.
%\eea
These expressions are consistent with the existing results of Ref.~[\onlinecite{xiao_arxiv_2014}] for 3d masless Dirac fermions.  
Further in the static limit {\it i.e.}, ${\tilde \omega}\to 0$, the imaginary part of the response function vanishes and 
Eqs. (\ref{3drepi0})- (\ref{3drepi1}) reduces to 
\be \label{static_gapless_3d}
\Pi_{0}(q,{\tilde \omega}=0)=-\frac{gq^2}{12\pi^2 \hbar v_{\rm F}} \log \left(\frac{2\Lambda}{q}\right)~,
\ee
and 
\bea \label{static_gapless_3d1}
\Pi_{1}(q,{\tilde \omega}=0) &=&\frac{-gq^2}{8\pi^2 \hbar v_{\rm F}}\Bigg[\frac{8\tilde{\mu}^2}{3q^2} -\frac{1}{6q^3}\Bigg\{ 2q^3 \log \left|\frac{4\tilde{\mu}^2 -q^2}{q^2}\right| \nn\\
& +& (8\tilde{\mu}^3 - 6\tilde{\mu} q^2) \log \left|\frac{2\tilde{\mu}-q}{2\tilde{\mu}+q}\right|\Bigg\}\Bigg]~.
\eea
Note that unlike the massive case, in Eq.~\eqref{static_gapless_3d1} the third derivative of the polarization function has a discontinuity at $q=2 k_{\rm F}$, and this leads to a $r^{-4}$ decay in the Friedel oscillations, as mentioned earlier and depicted in Fig.~\ref{fig4}(f) via the blue curve. 
The long-wavelength plasmon dispersion for the gapless $3$d Dirac system can be obtained by taking the ${\tilde \Delta} \to 0$ limit, in Eq.~(\ref{3dlw}).

\section{The non-relativistic limit of the massive Dirac polarization function}
\label{NR}
{ Note that while we have focussed on the polarization function for relativistic systems ($\Delta \to 0$, and finite $\Delta$) in this paper, the opposite limit of a dominant mass term is also very interesting. In the $\Delta \to \infty$ limit, the relativistic electronic dispersion reduces to a parabolic dispersion relation to leading order in the wave vector with $m \to \Delta/v_F^2$: 
\begin{equation}
E_{\bf k} (\Delta \to \infty) \approx \Delta + \frac{\hbar^2 k^2}{2 (\Delta/v_{\rm F}^2)}~.
\end{equation}
Thus for $\Delta \to \infty$, the long wavelength limit of the relativistic polarization function should reduce to the long wavelength limit of the non-relativistic polarization function in all dimensions. This is indeed the case, as can be inferred from Eq.~(10) of Ref.~[\onlinecite{rashi}], according to which the long wavelength limit of the polarization function for massive and massless Dirac systems, in all dimensions, is given by 
\be \label{Pilong}
\Pi^{(2d)}({\bf q},\omega) \approx  \frac{n_d}{\varepsilon_F/v_F^2} \frac{q^2}{\omega^2}~, 
\ee
where $n_d$ denotes the electron density in a $d$ dimensional spherically symmetric system. Now in the $\Delta \to \infty $ limit we can substitute $\varepsilon_{\rm F} \to \Delta$ and Eq.~\eqref{Pilong} just reduces to the long wavelength expression of the polarization function for a parabolic system.}

\section{Summary and Conclusions}\label{summary} 

To summarize, we have calculated the exact one-loop polarization function for massive as well as massless Dirac systems in 1d, 2d and 3d. {  The calculated polarization function is then used to obtain the exact plasmon dispersion, and the asymptotic form of the screened potential of a localized impurity. Both of these cannot be deduced from the long wavelength limit of the polarization function. The exact plasmon dispersion deviates from the long wavelength limit result at larger $q$, and it enters the p-h continuum generally at a larger $q$.  Additionally using the exact polarization function in the static limit, we find  that 
the Friedel oscillations in a massive Dirac system decay as $r^{-2}$ and $r^{-3}$ in 2d and 3d respectively similar to the case of parabolic dispersion and unlike the massless Dirac case where the corresponding Firedel oscillations decay as $r^{-3}$ and $r^{-4}$.}

For the case of $1$d massless Dirac fermions, the p-h continuum exists only along the $\omega = v_F q$ line, and additionally there are no Friedel oscillations. Both of these are a consequence of the fact that perfect backscattering or mixing of different chirality fermions in one dimension is forbidden. Another well known consequence of this forbidden backscattering or conservation of chirality is  Klein tunneling [\onlinecite{KT}] in massless Dirac systems such as graphene. 
For non-interacting massive Dirac fermions in $1$d, the density profile in vicinity of a static scatterer decays as $1/r$. However for interacting $1$d massive electron liquid, RPA fails to correctly describe the Friedel oscillations, which is a finite wave-vector phenomena. The correct behaviour of the Friedel oscillations as predicted by Luttinger liquid theory decays as $r^{-\alpha}$ with $\alpha <1$ for repulsive electron-electron interactions. %In $2$d and $3$d, Friedel oscillation in the screened potential decays as $1/r^2$ and $1/r^3$ respectively, similar to the case of parabolic fermions. However, for massless Dirac fermions, the decay rate is $1/r^3$ and $1/r^4$ in $2$d and $3$d respectively.

For the plasmon dispersion in massive Dirac systems, calculated both numerically and analytically in the long wavelength limit, we find that while the long wavelength limit behaviour in terms of $q$-dependence is purely determined by the charge continuity equation and dimensionality of the system as expected, the density dependence of the plasmon dispersion is governed by the details of the dispersion, doping etc.

We hope that our analytical results for the polarization function, plasmon dispersion and Friedel oscillations,  will be useful for exploring the physics of massive and massless Dirac electrons in different experimental systems with varying dimensionality. 

{  A useful extension of our work will be to include finite temperature effects in the calculation of the polarization function. A similar study has been done for the case of massless [\onlinecite{Qli, ramez}] and massive Dirac systems [\onlinecite{patel1}] in 2d. In general we expect the Firedel oscillations to be smeared out depending on the temperature (as compared to Fermi energy), and the plasmon dispersion to change quantitatively while retaining all the qualitative features.}

\section*{Acknowledgements}  We thank T. K. Ghosh for stimulating discussions. 
A.A. gratefully acknowledges funding from the INSPIRE Faculty Award by DST  (Govt. of India). We sincerely thank one of the PRB referees for independently checking and correcting several of the results in this manuscript. 
\vspace{0.5cm}
\appendix
\section{Calculation of Friedel Oscillations}
\label{FO}
The homogeneous or the long wavelength, $q \to 0$, behaviour of the static dielectric function, gives rise to the Thomas-Fermi screening which beautifully explains how the singularities of the long range Coulomb repulsion are `regularized' by screening. However 
Thomas-Fermi screening fails to adequately describe the response of the electron gas, to short range perturbations, for which the non-analyticity of the static dielectric function at finite wave-vector, typically at $q = 2 k_{\rm F}$, need to be accounted for [\onlinecite{Smbadlayan, Chenraikh, SimionGiuliani}].
To obtain the asymptotic behaviour of the screened potential from Eq.~\eqref{fdpotential} we follow Ref.~[\onlinecite{Smbadlayan}] and make use of the {\it Riemann-Lebesgue} lemma, which states that  if a function oscillates rapidly around zero then its integral is small and the principal contribution to the integral is determined by the behaviour of the integrand in the vicinity of its 
non-analytic points. In Eq.~\eqref{fdpotential}, since $\epsilon(q,0)$ is non-analytic at $q=2 k_{\rm F}$ in all three dimensions for Dirac systems, Eq.~\eqref{fdpotential} can be reduced to the following asymptotic form after doing the angular integration, 
\be \label{eq:potential2}
\phi(r) \approx A^{(d)} \int {\rm d}x~ \delta \Pi(x) S(x,r) e^{-x s}~,
\ee
where $\delta  \Pi(x) = \Pi(2 k_{\rm F}+x) - \Pi(2 k_{\rm F})$, is the increment of the static polarization function near the 
singular point $q=2 k_{\rm F}$ and $s \to 0$.  In Eq.~\eqref{eq:potential2}, we have defined $A^{(d)}$  to be a dimension dependent constant for  $d\ge1$ as
\be \label{A}
A^{(d)}= \frac{k_{\rm F}^{d-1} \phi_{\rm ext}(2 k_{\rm F}) V^{(d)}_{q=2k_{\rm F}} e^{-2k_{\rm F} s}}{[1-V^{(d)}_{q=2k_{\rm F}}\Pi^{(d)}(2k_{\rm F})]^2}~,
\ee
where $V^{(d)}_q$ is the Coulomb potential in $d$ dimensions in momentum space [see Eq.~\eqref{vq}],  and 
\bea
S(x,r)&=& 4 \pi \frac{\sin [(2k_{\rm F} + x)r]}{2 k_{\rm F} r}, \quad  ~~~~~~~~ d=3 \nn \\
&=& 2 \sqrt{2\pi}~ \frac{\cos[(2k_{\rm F}+x)r - \pi/4]}{\sqrt{2 k_{\rm F}r}}, \quad ~~~~~~~~d=2 \nn \\
&=& 2 \cos [(2k_{\rm F}+x)r ], \quad ~~~~~~~~d=1~.
\eea 
Note that in some cases where Eq.~\eqref{eq:potential2} cannot be solved analytically, its asymptotic behaviour for $r \to \infty$,  can be inferred simply from the lowest power of $x$ in the series expansion  of $\delta \Pi (x)$ around $x=0$. 

 As evident from Eq. (\ref{A}), $A^{(d)}$ depends on the value of static polarization function at the   $\omega = 0$ p-h boundary at $q=2k_{\rm F}$, and it vanishes if the static polarization function at $q=2 k_F$ diverges, as in the case of 1d systems (see Eq.~\eqref{eq:1dstatic1} for Dirac systems and Ref.~[\onlinecite{Giuliani_and_Vignale}] for parabolic systems). This is an example of the breakdown of  Fermi liquid theory and mean field  (RPA) based calculations to describe finite wave-vector phenomena in 1d. 
One dimensional interacting electron liquid, is more appropriately described by the Luttinger Liquid theory, and it has been shown that the asymptotic decay of the Friedel oscilations in 1d systems, has a power law decay where the decay exponent is dependent on the strength of the electron-electron interactions. Typically repulsive eletron-electron interactions slow down the decay of the Friedel oscillations as opposed to the non-interacting 1d case, where the Friedel oscillations decay as $r^{-1}$ [\onlinecite{Egger,Giuliani_and_Vignale}]. 

\end{document}